\input psfig
\documentstyle[12pt,aaspp4]{article}
\begin{document}
\title{Radiative Efficiency of Collisionless Accretion}
\author{Andrei V. Gruzinov}
\affil{Institute for Advanced Study, School of Natural Sciences,
Princeton, NJ 08540}

\begin{abstract}
Radiative efficiency, $\eta \equiv L/{\dot M}c^2$, of a slowly accreting black hole is estimated using a two-temperature model of accretion. The radiative efficiency depends on the magnetic field strength near the Schwarzschild radius. For weak  magnetic fields, $\beta ^{-1} \equiv B^2/8\pi p\lesssim 10^{-3}$, the low efficiency $\eta \sim 10^{-4}$ assumed in some theoretical models is achieved. For $\beta ^{-1} > 10^{-3}$, a significant fraction of viscous heat is dissipated by electrons and radiated away resulting in $\eta > 10^{-4}$. At equipartition magnetic fields, $\beta ^{-1} \sim 1$, we estimate $\eta \sim 10^{-1}$.
\end{abstract}
\eject

\section{Ion Tori, Advection Dominated Accretion Flows}

We will show that the radiative efficiency of an accreting black hole ($\eta \equiv L/{\dot M}c^2$, $L$ is the luminosity, ${\dot M}$ is the mass accretion rate) depends on the magnetic field strength. An investigation similar to ours, but with different conclusions, was carried out by Quataert (1997).

It was suggested that at low accretion rates, ${\dot M}\ll L_{Edd}/c^2$, a two-temperature plasma with $T_i\gg T_e$ forms near the Schwarzschild radius of an accreting black hole (Shapiro et al 1976, Phinney 1981, Rees et al 1982, Narayan \& Yi 1995), and the disk is inflated into a torus by the ion pressure. The plasma in the torus is collisionless, and Coulomb energy exchange between electrons and ions is small. Viscous heating power must be $\sim 0.1{\dot M} c^2$. If nearly all the heat is released into ions, radiative efficiencies as low as $\eta \sim 10^{-4}$ are obtained, because ions take all the heat into the hole. Electrons radiate efficiently, and $\eta$ can not be much smaller than $10^{-4}$ due to adiabatic heating of electrons. Observational estimates for $L$ and ${\dot M}$ exist for a number of objects ( Narayan et al 1997, Fabian \& Rees 1995 ), and some black hole candidates seem to have $\eta \lesssim 10^{-4}$. 

We show here that if indeed $\eta \lesssim 10^{-4}$, magnetic fields in these accretion flows must be very weak, $\beta ^{-1} \equiv B^2/8\pi p\lesssim 10^{-3}$, where $p$ is the gas (ion) pressure. For stronger magnetic fields, $\eta > 10^{-4}$ because electrons receive more than 0.1\% of the viscous heat, and this part of the viscous heat is efficiently radiated. At equipartition fields ($\beta ^{-1}\sim 1$), electrons get most of the heat and the radiative efficiency is close to the standard thin disk estimate , $\eta \sim 0.1$. Theoretical models of black hole accretion that assume both $\beta ^{-1} \sim 1$ and $\eta \lesssim 10^{-4}$ should be abandoned. 

The reason why stronger magnetic fields lead to an increased fraction of viscous energy going into electrons (and correspondingly to larger radiative efficiency) is roughly as follows. At strong fields, turbulence in the torus is dominated by Alfven waves which are only weakly damped at the length scales $\sim \rho$ (the ion Larmor radius), where damping by ions is maximal. Turbulent energy flux gets through the ion damping barrier at scales $\sim \rho$ and cascades to smaller scales. At length scales $\ll \rho$, ions drop out of dynamics because of the Larmor circle averaging, and the energy flux heats electrons.

The constraint $\beta ^{-1}\lesssim 10^{-3}$ (for $\eta \lesssim 10^{-4}$) is of interest for modeling the spectrum of an accreting black hole. The radiation spectrum is determined by the synchrotron emission and its Comptonization; the synchrotron peak frequency $\sim eB/mc$ changes by a factor of 30 when our estimate of $\beta ^{-1}$ is changed from 1 (which is the standard assumption) to $10^{-3}$.\footnote{The peak frequency also depends on the electron temperature. This dependence is non-trivial because the synchrotron emission is self-absorbed and comes from a range of distances from the hole.}

\section{Electron and Ion Heating}

Unless the plasma somehow leaves precisely all of its angular momentum behind by the time it reaches the hole, the ion torus will be rotating at about the Keplerian speed. To accrete, the plasma has to transport its angular momentum outwards (in a frame moving radially with the plasma), and this means that hydromagnetic turbulence should develop in the torus. Whatever the nature of turbulence and the resulting turbulent viscosity are, viscous heat will be generated at a rate $\sim 0.1{\dot M}c^2$ (e.g., Shapiro \& Teukolsky 1983). The question is into which plasma species is this heat deposited. The ions will take their share into the hole, while most of the electrons' share will be radiated. Narayan et al (1997) and Mahadevan (1997) suggested that the ratio of electron to ion heating, $\delta$, is given by the mass ratio, $\delta \sim m_e/m_i\sim 10^{-3}$. We will show that such small ratios of electron to ion heating are possible only at very weak magnetic fields.

There are two particle heating mechanisms in a magnetized collisionless plasma - Landau heating by electric fields parallel to the local magnetic field and Landau heating by the time varying magnetic field (by the curl of the electric field parallel to the local magnetic field, e.g. Stix (1992)). In the advection dominated regime (Narayan \& Yi 1995) near the hole, thermal velocities of electrons and ions are $v_e\sim v_i\sim c$, yet the $\gamma$-factors are not large, so that the plasma is roughly non-relativistic. At comparable thermal velocities Landau heating by parallel electric fields is proportional to $Q\sim e^2/m$, and hence it is dominated by electrons, the ratio of electron to ion heating being $\delta \sim m_i/m_e$. For magnetic heating, the charge $e$ should be replaced by the magnetic moment, $Q\sim \mu ^2/m$, where $\mu =mv^2/2B$. The heating ratio is $\delta \sim m_e/m_i$, as Narayan and coworkers usually assume. Electron to ion heating ratio will be much greater than $10^{-3}$ if heating by the parallel electric fields is important. Ion heating is also suppressed, and correspondingly electron to ion heating ratio is increased, at small length scales when averaging of the Lorentz force over the ion Larmor circle becomes important. 

To calculate the ratio of electron to ion heating, we must follow the turbulent energy as it cascades from the largest energy containing scales to smaller scales and estimate the strengths of the two damping mechanisms  for all scales. Since random magnetic fields limit the mean free path of plasma species to corresponding Larmor radii, hydrodynamic description is applicable at large scales, and Kolmogorov energy cascade develops. At smaller scales, turbulence is dominated by Alfven waves, because the nonlinear frequency $(\epsilon k^2)^{1/3}$ ($\epsilon$ is the energy flux in momentum space, units ${\rm cm}^2/{\rm s}^3$) becomes smaller than the linear Alfven frequency $v_Ak$ ($v_A$ is the Alfven speed). Following Goldreich \& Sridhar (1995) (Appendix A), we assume that for $k>\epsilon/v_A^3$, when Alfven frequency is formally larger than the nonlinear frequency, turbulence develops anisotropy, with filamentary perturbations aligned with the local magnetic field, $k_{\perp}\gg k_{\parallel}$. Degree of anisotropy is determined by a requirement that the spectrum be critically balanced, meaning that hydrodynamic and Alfven frequencies are comparable , $(\epsilon k_{\perp}^2)^{1/3}\sim v_Ak_{\parallel}$.

In the critically balanced Alfven turbulence regime, it takes about $t\sim 1/v_Ak_{\parallel}$ for a turbulent eddy to cascade to twice its wavenumber. Let $\gamma _k$ be the Landau damping rate at a given scale (calculated in Appendix B). The fraction of energy flux intercepted at wavenumber scales $k$ is of order $\gamma _k/v_Ak_{\parallel}$ - the rest of energy goes to scales of order $\sim 2k$.  The ratio $\gamma /\omega $ (damping to real frequency for linear waves) is all we need. This ratio determines at what scales the energy flux is damped, and knowing this it is straightforward to predict which plasma species is heated. Figure 1 shows $\gamma /\omega $ as a function of the perpendicular wavenumber. As seen from the figure, significant fraction of the energy is damped near the ion Larmor radius, $k_{\perp}\rho \sim 1$, when magnetic fields are weak, $\beta ^{-1}< 0.1$

For $k_{\perp}\rho \lesssim 1$ the main damping mechanism of Alfven waves is magnetic heating of ions. The damping rate is given by (Appendix B, for $\beta ^{-1}\ll 1$)
\begin{equation}
{\gamma \over \omega}={9\over 32{\sqrt \pi}} {v_i\over v_A}(k_{\perp}\rho )^2.
\end{equation}
At $k_{\perp}\rho \gtrsim 1$, Alfven waves gradually modify into whistlers (Appendices B and C), the ions still take most of the heat load, and for $\beta ^{-1}\ll 1$, 
\begin{equation}
{\gamma \over \omega}={1\over 2(k_{\perp}\rho )^2}.
\end{equation}
According to the last formula, ion damping becomes small at $k_{\perp}\rho \sim$ few. Turbulent energy that gets through the ion damping barrier at $k_{\perp}\rho \sim 1$ is only weakly damped until it reaches very small scales to be damped by electrons (see Appendix C for a discussion of whistler turbulence). Ions drop out of the picture at $k_{\perp}\rho \gtrsim$ few because of Larmor circle averaging. The ion heating maximum at $k_{\perp}\rho \sim 1$ determines the ratio of electron to ion heating. 

At equipartition magnetic fields ($\beta ^{-1}=1$), the maximal value of ion damping is $\gamma /\omega \approx 0.02$ at $k_{\perp}\rho \sim 1$ and much smaller at other scales. Most of the viscous heat gets over the ion damping barrier and is ultimately damped in electrons. Figure 2 shows the turbulent energy flow in the wavenumber space (arrows are drawn by hand!). The corresponding radiative efficiency is $\eta \sim 0.1$, similar to thin accretion disk models. 

The very small radiative efficiency assumed in advection dominated accretion models , $\eta \sim 10^{-4}$, is possible only if a very small fraction, $\delta \lesssim 10^{-3}$, of the Alfven wave energy at $k_{\perp}\rho < 1$  is transmitted to whistlers at $k_{\perp}\rho \sim$ few. Whistlers can propagate at $k_{\perp}\rho$ as small as $\sim 1$. For $\beta > 10^{-3}$, Alfven waves are not fully damped at $k_{\perp}\rho \approx 0.5$ and will excite whistlers by three-wave interactions. Thus, $\beta ^{-1}\lesssim 10^{-3}$ is necessary for $\eta \lesssim 10^{-4}$.

Fortunately, electron parameters ($v_e$ and $m_e$) do not enter Eqs. (1) and (2). Electron thermal velocity and $\gamma$-factor are determined by radiation and are not reliably predicted by the advection dominated (or any other) accretion model. However, the only parameter that matters for our calculation is $\beta ^{-1}$ - an input parameter. Our predictions are not limited to any particular model of accretion, as long as $T_i\gg T_e$ and $v_A < v_i < v_e$.

\section{Discussion}

A minimal model of turbulence was assumed in this work - no secondary instabilities, no shocks, just a simple Kolmogorov cascade with Landau damping. This model predicts substantial heating of electrons and large radiative efficiencies if magnetic fields are close to equipartition. Small radiative efficiency is possible if nearly all the viscous heat is absorbed by ions. For this, magnetic fields must be very weak, $\beta ^{-1}\lesssim 10^{-3}$. 

Incidentally, we can argue that low radiative efficiency is possible only at very weak magnetic fields if we assume that Blandford \& Znajek (1977) mechanism is at work. Poloidal magnetic field of order $B_p$ near a maximally rotating black hole gives a luminosity $L\sim B_p^2R_g^2c$. We get a radiative efficiency $\eta \sim (B_p/B)^2\alpha ^{-1}\beta ^{-1}$, where $\alpha$ is the Shakura-Sunayev parameter. Arbitrarily assuming $B_p/B\sim 0.1$, $\alpha \sim 0.1$, we get $\eta _{BZ}\sim 0.1\beta ^{-1}$, so that $\beta ^{-1}<10^{-3}$ is required for $\eta _{BZ}<10^{-4}$. 

\acknowledgements

I thank John Bahcall and Martin Rees for suggesting the problem, and Peter Goldreich, Pawan Kumar and Insu Yi for valuable discussions. This work was supported by NSF PHY-9513835.

\appendix

\section{Strong Turbulence of Alfven Waves According to Goldreich \& Sridhar}

This Appendix is based on Goldreich \& Sridhar (1995). Suppose that $\epsilon$ is the energy flux in the wavenumber space (energy pumping rate, units ${\rm cm}^2/{\rm s}^3$). Isotropic Kolmogorov turbulence is characterized by a nonlinear frequency $\sim (\epsilon k^2)^{1/3}$ at wavenumbers $\sim k$. If there is a large scale magnetic field with Alfven speed $v_A$, another characteristic frequency of the turbulence is the linear frequency $v_Ak_{\parallel}$. If the turbulence were isotropic, the hydrodynamic nonlinear frequency would be much smaller than the linear Alfven frequency at small enough scales. According to Goldreich \& Sridhar this does not happen. Instead, turbulent energy propagates to smaller scales within a narrow cone in the $k$-space given by $k_{\parallel}\lesssim (\epsilon ^{1/3}/v_A)k_{\perp}^{2/3}$. In other words, turbulence is dominated by elongated structures, that are aligned with the local magnetic field. This could be roughly explained as follows.

Suppose, that a ``turbulent eddy'' leaves the cone. On the outside of the cone, turbulence is just a collection of weakly nonlinear waves. Waves propagating parallel and antiparallel to the local magnetic field follow the field lines perturbed by each other, and thus cascade in $k_{\perp}$. Turbulent energy from outside of the cone will cascade in $k_{\perp}$ without cascading in $k_{\parallel}$ and will thus enter the cone.

Turbulence fills the entire cone, because within the cone the presence of the large-scale magnetic field is not ``felt'', $v_Ak_{\parallel}<(\epsilon k^2)^{1/3}$. Turbulent eddies will try to isotropize themselves, filling as much of the $k$-space as possible. 

\section{Landau Damping of Alfven Waves and Whistlers}

We calculated Landau damping of Alfven waves (whistlers at $k_{\perp}\rho \gtrsim 1$) numerically using a full dielectric permittivity tensor of a magnetized Maxwellian plasma (e.g. Mikhailovskii 1967). Results are shown in Fig.1, where the ratio of imaginary to real parts of the eigenfrequency is plotted versus dimensionless perpendicular wavenumber $k_{\perp}\rho$. The calculation was carried out for Maxwellian electrons and ions with thermal velocities $v_e/c=0.66$ and $v_i/c=0.33$ which are close to the plasma parameters of a two-temperature advection dominated accretion model of Narayan \& Yi (1995). We assumed $k_{\parallel}/k_{\perp}=10^{-2}$, and checked that $\gamma /\omega$ does not depend on $k_{\parallel}/k_{\perp}$ as long as $k_{\parallel}/k_{\perp}\ll 1$. 

The four curves correspond to four values of $\beta ^{-1}$. Damping by ions is maximal at $k_{\perp}\rho \sim 1$, where damping by electrons is negligible. For strong fields, $\beta ^{-1}\sim 1$, the damping at $k_{\perp}\rho \sim 1$ is small. The damping increases with decreasing $\beta ^{-1}$. At $\beta ^{-1}\sim 10^{-2}$, the damping is so strong that Alfven waves do not propagate  in a narrow region near $k_{\perp}\rho \approx 0.85$. The region of forbidden propagation is narrow for $\beta ^{-1}\sim 10^{-2}$ and widens for $\beta ^{-1}\sim 10^{-3}$. 

Two remarks are in order. Electrons and ions that resonate with Alfven waves and whistlers belong to the bulk of the distribution functions. Deviations from Maxwellian distribution, which may be expected in our collisionless plasma, will not strongly affect the results. Also, linear and nonlinear Landau damping rates are similar, because linear frequencies and nonlinear frequency shifts are of the same order for a critically balanced turbulence.

Most of the numerical results shown in Fig.1 can be understood analytically. As discussed in \S 2, analytical expressions (1) and (2) are of importance - they imply that our results are model independent. For $k_{\perp}\gg k_{\parallel}$, $v_A\lesssim v_i\lesssim v_e\lesssim c$, the following approximate expressions for the permittivity tensor (Mikhailovskii 1967) give a good accuracy. For a wavenumber in the $xz$-plane, with $z$ along the local magnetic field,
\begin{eqnarray}
\epsilon _{xx} & \equiv & \epsilon _1={c^2\over v_A^2}{1-I_0e^{-z}\over z},\\
\epsilon _{xy} & \equiv & ig=i{c^2\over v_A^2}{v_i\over v}{k_{\perp}\over k_{\parallel}}{1\over k_{\perp}\rho}\{ 1-(I_0-I_1)e^{-z}\} ,\\
\epsilon _{xz} & =      & 0,\\
\epsilon _{yy} & \equiv & \epsilon _2=i{\sqrt \pi}W(v/v_i)({k_{\perp}\over k_{\parallel}})^2{c^2\over v_A^2}{v_i\over v}(I_0-I_1)e^{-z},\\
\epsilon _{yz} & =      &0,\\
\epsilon _{zz} & \equiv & \epsilon _3=2{m_i\over m_e}{c^2v_i^2\over v_A^2v_e^2}({k_{\perp}\over k_{\parallel}})^2{1\over (k_{\perp}\rho )^2}\{ 1+i{\sqrt \pi}{v\over v_e}W(v/v_e)\} .
\end{eqnarray}
Here $v=\omega /k_{\parallel}$ is the phase speed, $z=(k_{\perp}\rho )^2/2$, $I=I(z)$ are modified Bessel functions, and
\begin{equation}
W(x)=e^{-x^2}(1+{2i\over {\sqrt \pi}}\int_0^xdte^{t^2}).
\end{equation}
Permittivities $\epsilon _1$ and $\epsilon _2$ are due to ions, $\epsilon _3$ is due to electrons, and $g$ is due to both species. 

The dispersion relation for linear waves follows from Maxwell equations,
\begin{equation}
\det (\epsilon _{ij}+{c^2\over \omega ^2}(k_ik_j-k^2\delta _{ij}))=0.
\end{equation}
For $k_{\perp}\rho \lesssim 10$, the parallel permittivity $\epsilon _3$ is very large, and both damping and real frequency are accurately predicted by a two dimensional approximation 
\begin{equation}
(\epsilon _1-{c^2k_{\parallel}^2\over \omega ^2})(\epsilon _2-{c^2k^2\over \omega ^2})-g^2=0.
\end{equation}

For $k_{\perp}\rho \ll 1$, we have $\epsilon _1\epsilon_2\gg g^2$, and Eq. (B9) reduces to $\epsilon _1=c^2/v^2$, which gives the Alfven wave dispersion law  
\begin{equation}
v=v_A.
\end{equation} 
For $k_{\perp}\rho \gg 1$, we have $c^2k_{\parallel}^2/\omega ^2\gg \epsilon _1$, $c^2k^2/\omega ^2\gg \epsilon _2$, and Eq. (B9) gives $c^2k_{\parallel}k/\omega ^2=g$, that is whistlers 
\begin{equation}
v=(v_A^2/v_i)\rho k. 
\end{equation}

Corrections to the above dispersion laws (within the 2D approximation, Eq. (B9)) give the damping rates for Alfven waves and whistlers. For Alfven waves, in the limit $v_A\ll v_i$ when $\epsilon _2$ is pure imaginary, we have approximately
\begin{equation}
(\epsilon _1-{c^2k_{\parallel}^2\over \omega ^2})\epsilon _2=g^2,
\end{equation}
giving
\begin{equation}
{\gamma \over \omega }={g^2\over 2\epsilon _1\epsilon _2},
\end{equation}
that is 
\begin{equation}
{\gamma \over \omega}={9\over 32{\sqrt \pi}} {v_i\over v_A}(k_{\perp}\rho )^2.
\end{equation}
For whistlers at $k_{\perp}\rho \gtrsim 1$, we have approximately 
\begin{equation}
{c^2k_{\parallel}^2\over \omega ^2}({c^2k^2\over \omega ^2}-\epsilon _2)=g^2,
\end{equation}
giving
\begin{equation}
{\gamma \over \omega}={1\over 2(k_{\perp}\rho )^2}.
\end{equation}

At even larger $k_{\perp}\rho$ electron damping becomes important. The ratio of electron to ion damping can be estimated (for $v_A\ll v_i$) as
\begin{equation}
\delta = {\epsilon _3^{im}|E_z|^2\over \epsilon _2^{im}|E_y|^2}.
\end{equation}
Here the wave electric fields should be obtained from the system of linear equations whose determinant is given in Eq. (B8). Two of these equations can be approximately written as
\begin{equation}
{c^2k_{\parallel}k_{\perp}\over \omega ^2}E_x=\epsilon _3E_z,
\end{equation}
\begin{equation}
{c^2k_{\perp}^2\over \omega ^2}E_y=igE_x.
\end{equation}
Now Eq. (B17) gives
\begin{equation}
\delta = {{\sqrt \pi }\over 2}{m_ev_e\over m_iv_i}(k_{\perp}\rho )^3,
\end{equation}
showing that electron heating dominates for $k_{\perp}\rho \gtrsim 10$. At these small scales, ions are not important either for real part of the frequency or its imaginary part (heating). For perturbations with perpendicular length scales $\lesssim \rho /10$, ions can be treated as a constant positive background for electrons and electromagnetic fields. 

\section{EHD and Whistler Turbulence}

We should make sure that turbulent energy will keep cascading to smaller scales after Alfven waves are transformed into whistlers at length scales $\sim \rho$. Like collisionless Alfven waves, collisionless whistlers can be described hydrodynamically. Hydrodynamic equations valid at length scales $\lesssim \rho$ can be obtained as follows. Electrons move freely along magnetic field lines, and $E_{\parallel}=0$.  In the perpendicular direction, electrons move with the E$\times $B-drift velocity, ${\bf v}_{\perp }=c{\bf E}\times {\bf B}/B^2$. These two equations give 
\begin{equation}
{\bf E}+{1\over c}{\bf v}\times {\bf B}=0.
\end{equation}
Neglecting displacement currents, 
\begin{equation}
\nabla \times {\bf B}=-{4\pi \over c}ne{\bf v}.
\end{equation}
From Eqs. (C1) and (C2), we obtain the Electron Hydrodynamics (EHD) equation (Kingsep, Chukbar, \& Yan'kov 1990)
\begin{equation}
\partial _t {\bf B}={c\over 4\pi ne}\nabla \times ({\bf B}\times \nabla \times {\bf B}).
\end{equation}
Linear waves in Eq. (C3) are whistlers, their dispersion law can be presented as Eq. (B11). Turbulent cascade in Eq. (C3) was discussed by Kingsep, Chukbar, \& Yan'kov (1990) and by Goldreich \& Reisenegger (1992).

\begin{figure}[htb]
\psfig{figure=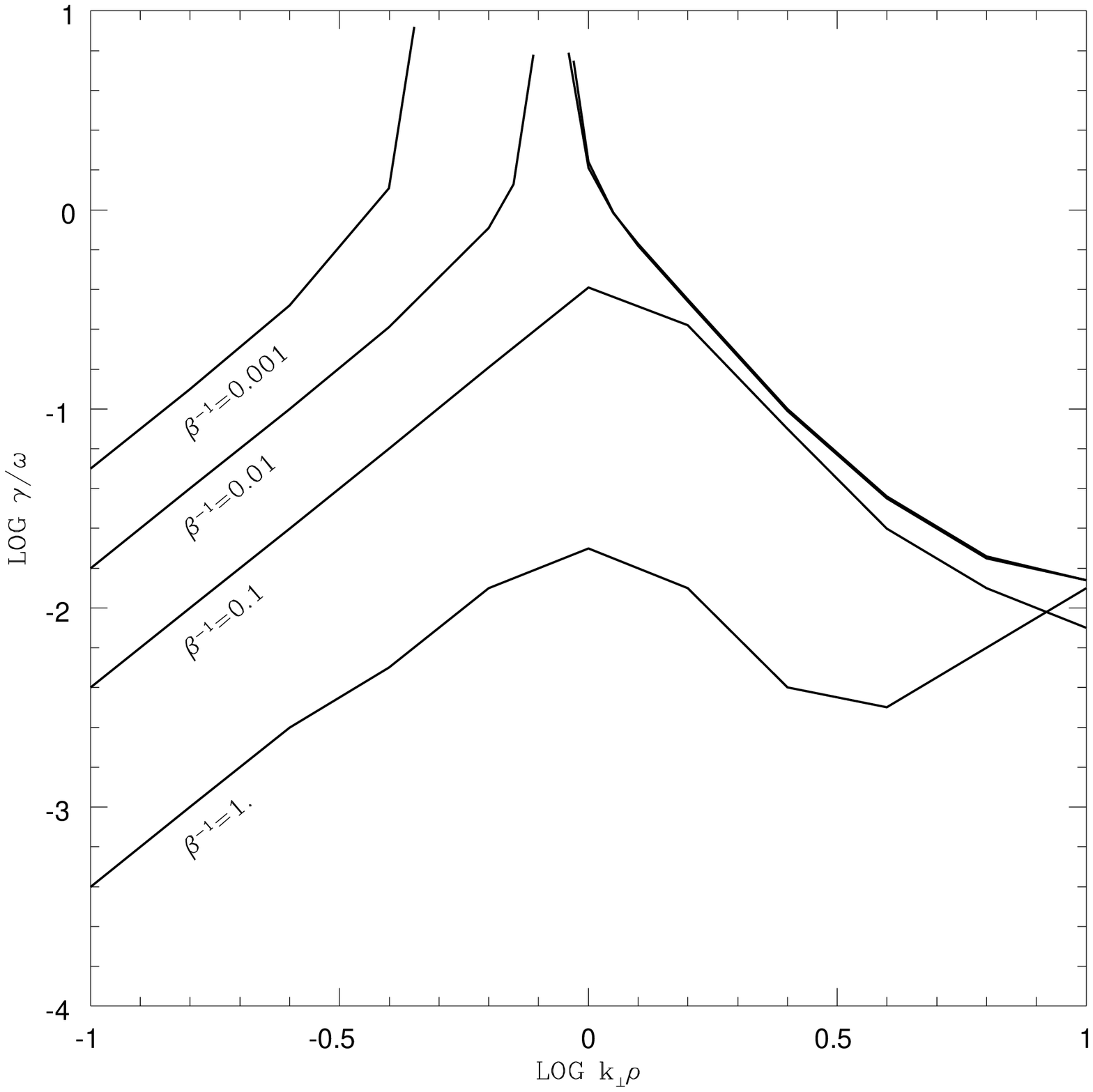,width=7in}
\caption{Damping rate over real frequency versus $k_{\perp}\rho$ for plasma parameters given in Appendix B. For $k_{\perp}\rho <10$, ions are taking most of the heat, with the exception of the $\beta ^{-1}=1$ case; the rise of the curve at $\log k_{\perp}\rho >0.5$ is due to electron heating.}
\end{figure}

\begin{figure}[htb]
\psfig{figure=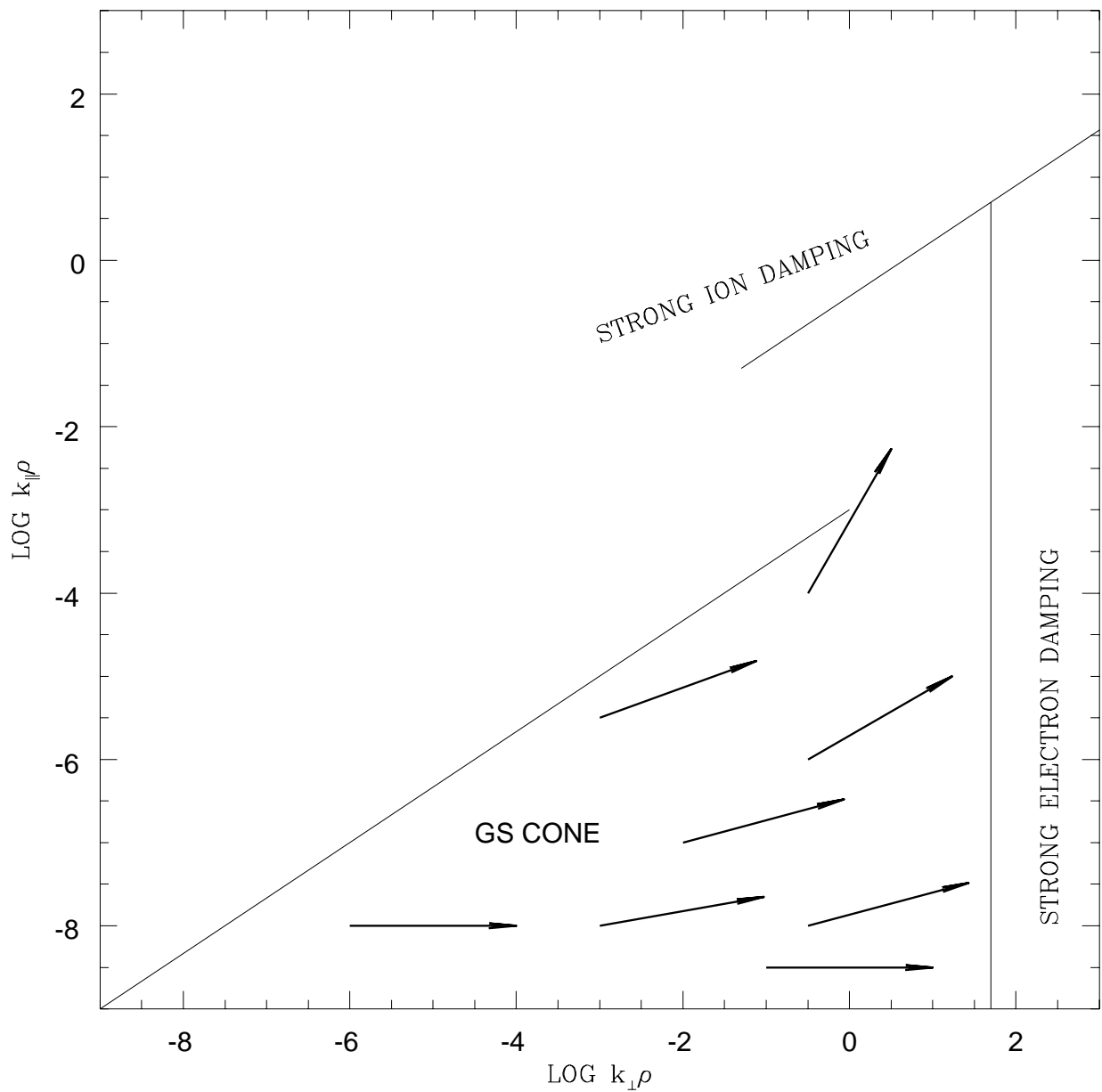,width=7in}
\caption{Energy flow in the wavenumber space for $\beta ^{-1}=1$. Narayan and Yi (1995) formulae were used to estimate the parameters. In their notation, black hole mass $m\sim 10^6$, Shakura-Sunyaev parameter $\alpha \sim 0.1$, mass accretion rate in Eddington units ${\dot m}\sim 0.001$, turbulent energy-containing length scale in units of the Schwarzschild radius $L\sim 0.3$, large scale turbulent velocity $\sim 0.1c$.}
\end{figure}

\end{document}